\documentclass[prl,twocolumn,superscriptaddress]{revtex4}
\usepackage{graphicx}
\usepackage{amsmath}
\usepackage{amssymb}

\begin{document}

\title{Evidence for thermal spin transfer torque}

\author{Haiming Yu}
  \affiliation{Ecole Polytechnique F\'{e}d\'{e}rale de Lausanne, IPMC, Station~3, CH-1015 Lausanne-EPFL, Switzerland}
  \affiliation{State Key Laboratory for Mesoscopic Physics, School of Physics, Peking University, Beijing 100871, People's Republic of China}
\author{S. Granville}
  \affiliation{Ecole Polytechnique F\'{e}d\'{e}rale de Lausanne, IPMC, Station~3, CH-1015 Lausanne-EPFL, Switzerland}
\author{D. P. Yu}
  \affiliation{State Key Laboratory for Mesoscopic Physics, School of Physics, Peking University, Beijing 100871, People's Republic of China}
\author{J.-Ph. Ansermet}
  \affiliation{Ecole Polytechnique F\'{e}d\'{e}rale de Lausanne, IPMC, Station~3, CH-1015 Lausanne-EPFL, Switzerland}

\begin{abstract}
Large heat currents are obtained in Co/Cu/Co spin valves positioned at the middle of Cu nanowires. The second harmonic voltage response to an applied current is used to investigate the effect of the heat current on the switching of the spin valves. Both the switching field and the magnitude of the voltage response are found to be dependent on the heat current. These effects are evidence for a thermal spin transfer torque acting on the magnetization and are accounted for by a thermodynamic model in which heat, charge and spin currents are linked by Onsager reciprocity relations.
\end{abstract}

\pacs{}

\date{\today}

\maketitle

It has been established for some years now that an electrical current may be used to induce magnetization reversal in nanostructures such as spin valves.  This current-induced magnetization switching (CIMS) operates by way of a current-induced spin transfer torque (STT) that acts on the magnetization of the switching layer. CIMS could be a technological alternative to magnetization reversal by application of an external magnetic field in the development of various devices based on spin valves such as high density and low power magnetic memories~\cite{Slonczewski}. Phenomena such as CIMS fall into the rapidly developing field of spintronics, whereby the spin as well as charge degrees of freedom are exploited simultaneously for information processing and storage.

The field of ``spin caloritronics'', i.e., the addition of thermal effects to the electrical and magnetic properties of nanostructures, has recently seen a surge in interest from both theoretical~\cite{Hatami_Bauer2} and experimental perspectives~\cite{Katayama,Gravier_Fukushima,Uchida}. It has most recently been predicted that a heat current can exert a torque on the magnetization in nanostructures such as spin valves~\cite{Hatami_Bauer1,Heikkila}, or on domain walls within a magnetic nanowire~\cite{Yuan,Bauer,Kovalev2,Kovalev1}.  These theoretical studies have explored a variety of novel effects, including domain-wall-motion-induced Peltier cooling or power generation, and devices such as nanoscale heat pumps or rotational nanomotors are envisaged.  Thermal spin transfer torque driven by a heat current is expected to be highly efficient: Yuan \textit{et al}.~\cite{Yuan} calculated that temperature differences of 1~K generate spin torques in Ni and Co domain walls two orders of magnitude larger than the STT obtained at the critical electrical current density for domain wall motion.  Bauer \textit{et al}.~\cite{Bauer} used parameters typical of permalloy to show that a temperature gradient of 0.2~K/nm is as efficient as a charge current density of 10$^{7}$~Acm$^{-2}$.  These calculations indicate the great potential of heat currents for driving domain-wall based devices or magnetic memories based on spin valves.

In this paper, we provide experimental evidence that a strong heat current can indeed affect the magnetization dynamics in nanostructures. Experiments are conducted on spin valves designed to have a heat current. Control structures without a heat current allow us to rule out possible spurious effects such as the overall temperature rise.

\begin{figure}
\centering{
  \includegraphics[width=8cm]{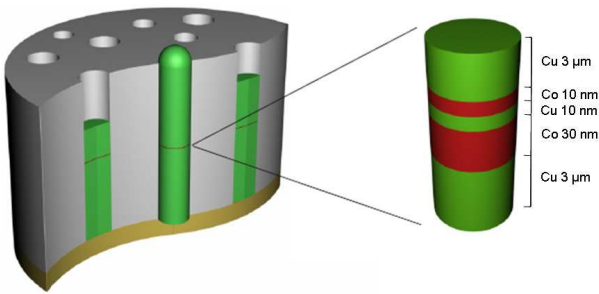}
} \caption{Cartoon schematic of a sample of one spin valve inside a Cu nanowire. The first few Cu nanowires that reach the top surface of the nanoporous membrane form overgrowths used as contact pads.
}
\label{fig1}
\end{figure}

Using electrodeposition in nanoporous membranes, we form Co/Cu/Co spin valves midway in Cu nanowires, with one Co layer much thicker than the other. Figure~\ref{fig1} is a cartoon schematic of the resulting sample. The top contact is obtained by rubbing a Au wire onto the top surface of the membrane.  Complete growth details, including the method of making contacts to single nanowires, are provided in a previous publication by our group (Ref.~\cite{Biziere}). The statistical distribution of the pore diameters has a standard deviation of about 10~nm and an average diameter of 50~nm~\cite{Chlebny_Doucin} (pore density $6\times10^{8}$~cm$^{-2}$). When a current is driven through the nanowire, the Joule effect generates heat in the more resistive Co layers.  The heat dissipated in the thicker layer flows through the thinner layer and out the Cu lead. The temperature profile can be obtained by integration of the Fourier equation in the stationary regime, with the massive electrodes at the ends of the nanowire remaining at a set temperature. The calculation yields a result close to that of the simple argument according to which the heat is assumed to flow equally in both halves of the nanowire.  In view of this, we have:

\begin{equation}\label{Joule}
\frac{1}{2}\rho\frac{d}{\pi r^{2}}I^{2}=j_{Q}\pi r^{2},
\end{equation}

\noindent where $\rho$ is the resistivity of electrodeposited Co taken from previous work~\cite{Gravier}, $d$ the thickness of the thick layer, and $\pi r^{2}$ the cross-sectional area of the nanowire. A large heat current $j_{Q}$~$\propto$~1/$r^4$ can be expected if the radius $r$ of the nanowire is small enough. With $j_{Q}=-\kappa\nabla T$ and taking for $\kappa$ a typical value of $10 \frac{W}{mK}$, a temperature gradient as large as 1000~K/cm is expected for a current of 100~$\mu A$.  We also produced symmetric spin valves with Co layers of equal thicknesses. By symmetry, the temperature gradient in the symmetric spin valves is zero and there is nearly no heat current.

Our measurement of the effect of heat current on the spin valves makes use of a technique we have recently demonstrated to be sensitive to spin torque effects in magnetic nanostructures~\cite{Yu_Dubois}. We apply simultaneously a dc current $I_{\textrm{dc}}$ and an ac current $I_{\textrm{ac}}$ with frequency $f$$\sim$400 Hz. We can expect the temperature increase in the nanostructure to be proportional to the square of the current. As the square of a sine wave contains a dc component and a component oscillating at twice the frequency, the current produces both a fixed increase and an oscillation of temperature in each layer of the spin valve.  We thus expect a dc heat current and a heat current oscillating at $2f$. The coupling of an ac current with the giant magnetoresistance (GMR) results in a dc voltage, the well-known spin diode effect~\cite{Yuasa,Ralph}, and also implies a voltage at $2f$ proportional to $I_{\textrm{dc}}$ (second term of Eq.~(\ref{V2f}) below). There is also a voltage at frequency $2f$ when $I_{\textrm{dc}}$ is set at zero; this is the linear response to spin transfer torque~\cite{Yu_Dubois}. We monitor the second harmonic voltage response $V^{2f}$ whilst sweeping an externally applied magnetic field. We use the result that a peak appears in the $V^{2f}$ signal at the magnetic field $H_{sw}$ where a Co layer magnetization switches. We found that the $V^{2f}$ has a higher signal-to-noise ratio than GMR, which allows for a cleaner view of the switching process.

\begin{figure}\vspace{-0.8 cm}
\centering{
  \includegraphics[width=8cm]{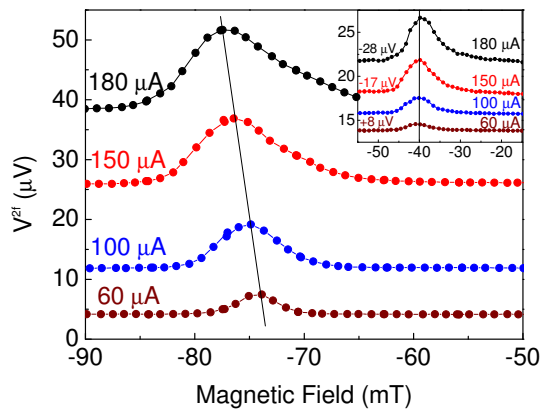}
}\vspace{-0.4 cm} \caption{$V^{2f}$ measurements on a nanowire containing an asymmetric spin valve (10~nm~Co/10~nm~Cu/30~nm~Co) at a set dc current of -100~$\mu$A for different values of the ac current, as indicated in $\mu$A. The field was swept from negative to positive values. Inset: idem for a symmetric spin valve (10~nm~Co/10~nm~Cu/10~nm~Co). Curves are offset vertically by the values specified in $\mu$V. The lines are guides for the eye.
}
\label{fig2}
\end{figure}

\begin{figure}\vspace{-0.8 cm}
\centering{
  \includegraphics[width=8cm]{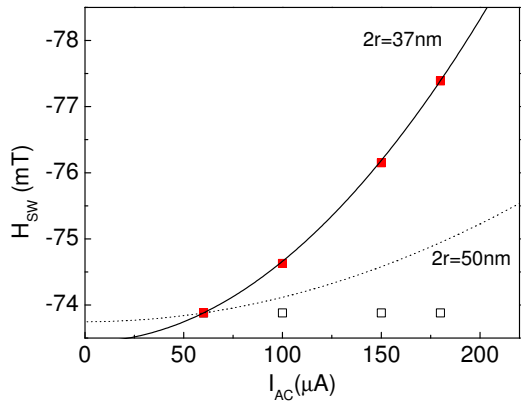}
} \vspace{-0.4 cm} \caption{Switching field measured as a function of $I_{\textrm{ac}}$, for fixed $I_{\textrm{dc}}$=-100~$\mu$A. Squares (red) are the experimental data for an asymmetric spin valve. The open boxes (black) are the data for a symmetric spin valve. The lines (black) are the result of calculations with different nanowire diameters as indicated in the text.
}
\label{fig3}
\end{figure}

For asymmetric spin valves (10Co/10Cu/30Co) (units in nm), we observe a clear dependence of $H_{sw}$ on the applied ac current (Figs.~\ref{fig2} and~\ref{fig3}), which we attribute to the effect of the dc heat current on the magnetization of the thin Co layer. This effect is not observed in symmetric spin valves (10Co/10Cu/10Co10) (Fig.~\ref{fig2} inset and Fig.~\ref{fig3}). In the following we demonstrate that our observations are due neither to an ac spin torque effect nor to a temperature rise of the sample.

For a symmetric spin valve, there appears to be no detectable effect of an ac spin torque on $H_{sw}$, as shown in the inset of Fig.~\ref{fig2} and in Fig.~\ref{fig3}. However, we find that $I_{\textrm{dc}}$ does change $H_{sw}$, confirming that a dc spin transfer torque affects the switching field. This can be understood because the magnetization switching process in our samples is reversible rather than hysteretic, that is, there are no minor loops observed using either GMR or $V^{2f}$ measurements.  The switch occurs over several mT as the applied field is swept, during which non-collinear configurations of the two Co layers occur in a quasi-static regime.  Consequently, the ac current only causes the magnetization to oscillate about the stationary configuration determined by the applied field and the dc current. Thus, only a dc torque can affect $H_{sw}$.  In the measurements of Figs.~\ref{fig2} and~\ref{fig3}, $I_{\textrm{dc}}$ is fixed, so the only possible source of additional dc torque is the dc heat current generated by the increasing $I_{ac}$.

When a current is driven through a nanowire, its temperature rises. We determine the temperature rise induced by ac currents of 100~$\mu A$ or 180~$\mu A$ to be about 0.8~K or 2.5~K respectively, using the observed rise of the nanowire resistance and a measurement of the temperature dependence of resistance using a lower current. Calculations of the temperature profile in a nanowire under Joule heating are consistent with these values. Therefore, we must consider the possible influence on $H_{sw}$ of this temperature increase. Two observations allow us to neglect this effect.

First, we made measurements of $H_{sw}$ using an external heat source to heat the nanowires, in which we see that a 30~K temperature increase causes a change of $H_{sw}$  of only 1.2~mT. Hence, with the 0.8~K temperature rise during the $V^{2f}$ measurements we expect an $H_{sw}$ shift of 0.03~mT.  This effect is insufficient to account for the actual $H_{sw}$ shift observed during the measurement of Fig.~\ref{fig2}, which is two orders of magnitude larger.

Second, an applied current generates almost the same overall temperature rise in a nanowire containing a symmetric spin valve as in one containing an asymmetric spin valve. Therefore, if the $H_{sw}$ shift in the asymmetric spin valve was caused by the temperature increase under $I_{\textrm{ac}}$, we should see about the same $H_{sw}$ shift in the symmetric spin valve, however, as mentioned before, $H_{sw}$ is constant with increasing $I_{\textrm{ac}}$. The key difference between these two samples is that in an asymmetric spin valve we have a large dc heat current passing through the thin layer whereas in a symmetric spin valve the equivalent Joule heating in each Co layer results in a zero net heat current passing through the spin valve. Thus, we conclude that in Fig.~\ref{fig2}, the $H_{sw}$ shift is due to the dc part of the heat current caused by Joule heating.

\begin{figure}\vspace{-0.8 cm}
\centering{
  \includegraphics[width=8cm]{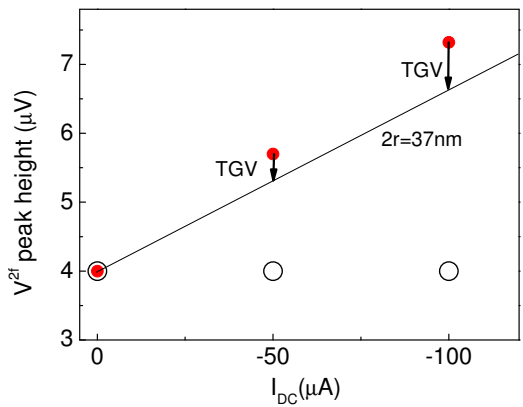}
}\vspace{-0.4 cm} \caption{$V^{2f}$ peak height as a function of $I_{\textrm{dc}}$, for fixed $I_{\textrm{ac}}$=100~$\mu$A. The dots (red) are the experimental data for an asymmetric spin valve. The open circles (black) are the experimental data for a symmetric spin valve. The arrows indicate a linear correction to the data of the asymmetric spin valve by subtraction of an estimated value for the TGV. The line (black) is a calculation using the nanowire diameter of 37~nm as indicated in the text.
}
\label{fig4}
\end{figure}

The peak height of $V^{2f}$ also shows the effect of a heat current.  Here we apply a fixed $I_{\textrm{ac}}$ and vary $I_{\textrm{dc}}$.  The data of Fig.~\ref{fig4} show that in a symmetric spin valve the peak height of $V^{2f}$ is independent of $I_{\textrm{dc}}$, but in an asymmetric spin valve $V^{2f}$ has a clear linear dependence on $I_{\textrm{dc}}$. By considering the various contributions to $V^{2f}$, we can determine the origin of this effect. We can write:

\begin{align}\label{V2f}
V^{2f}=&\frac{dR}{d\tau}\left(I_{\textrm{ac}}~\tau_{STT}^{f}+I_{\textrm{dc}}~\tau_{TST}^{2f}\right)+I_{\textrm{dc}}\frac{dR}{dT}\Delta T^{2f}.
\end{align}

\noindent The first term is due to an ac spin transfer torque, which is independent of $I_{\textrm{dc}}$~\cite{Yu_Dubois}. The second torque term contains a thermal spin torque $\tau_{TST}^{2f}$ that oscillates at frequency $2f$, because the heat current has a component at this frequency. This term is proportional to $I_{\textrm{dc}}$. The last term is simply the contribution of the temperature dependence of $R$.

We can understand the coupling of heat current and magnetization in the framework of a thermodynamic approach. The relevance of such an approach has been shown before, starting with the seminal work of Johnson and Silsbee~\cite{Johnson_Silsbee,Johnson_Byers,Wegrowe}. Because of the thickness of the layers in the range of 10 nm or above, and due to the defects that one can expect in electrodeposited metals, it is clear that electronic transport is in the diffusive regime in which the thermodynamic approach is relevant. The Onsager reciprocity relations imply linear relations between currents of heat, charge and spin on one side, and on the other side their associated generalized "forces", the gradients of temperature and of electrochemical potential. In this linear regime we may expect the torque to be proportional to the spin current $\mathbf{j_{m}}$, which has two major contributions, one from the gradient of the electrostatic potential $\nabla V$, the other from the gradient of temperature $\nabla T$. In our experiment, both gradients are imposed on the thermodynamic system. Using the notation of Ref.~\cite{Dubois_Ansermet}~(Eq. (18))

\begin{equation}\label{jm}
j_{m}=2c(\nabla V-S_{eff}\nabla T)
\end{equation}

\noindent with $S_{eff}
=\epsilon_{0}\left(1+\frac{\eta}{\beta}\right)$, which has the same units as a Seebeck coefficient, where $\epsilon_{0}$ is the Seebeck coefficient of Co, $\eta$ and $\beta$ are the conventional spin asymmetries of the Seebeck coefficient and resistivity respectively and $c$ is the spin-dependent part of the conductivity.

A torque \textbf{$\tau$} was calculated for a spin valve by Hatami \textit{et al}.~\cite{Hatami_Bauer1}, considering spin-dependent heat and charge transport at the interface between a ferromagnetic metal and a normal metal. Two contributions are expected, with some similarity to Eq.~(\ref{jm}):

\begin{equation}\label{TSTHatami}
\mathbf{\tau}\propto P\Delta V + P^{\prime}S\Delta T,
\end{equation}

\noindent where $P$ and $P^{\prime}$ characterize respectively the spin asymmetry of the conductivity and of the Seebeck coefficient $S$. However, as Hatami \textit{et al}. pointed out in their concluding remarks, their calculation concerns interface spin effects. Thus, it is a temperature difference $\Delta T$ which is relevant in their case. In the diffusive transport regime, it is the temperature gradient $\nabla T$ that plays a role: at 1000~K/cm this is quite large, whereas the temperature difference between the two layers is of the order of only 1~mK. In order to account for our data using Eq.~(\ref{TSTHatami}), we need to assume for $S$ a value 1000 times larger than typical values for the Seebeck coefficient in metals. We henceforth interpret the data using Eq.~(\ref{jm}).

The change of switching field due to the torque $\tau_{TST}^{0}$ associated with the dc heat current relative to that due to the spin transfer torque $\tau_{STT}^{0}$ is estimated by:

\begin{equation}\label{STratio}
\frac{\Delta H_{sw}^{TST}}{\Delta H_{sw}^{STT}}=\frac{\tau_{TST}^{0}}{\tau_{STT}^{0}}=\frac{j_{m,TST}^{0}}{j_{m,STT}^{0}}=\frac{S_{eff}\nabla T}{\nabla V}.
\end{equation}

\noindent As demonstrated by Fig.~\ref{fig2}, $\Delta H_{sw}^{STT}$ is independent of $I_{\textrm{ac}}$, and the value was determined experimentally with $I_{\textrm{dc}}=-100~\mu$A. Thus, for a fixed $I_{\textrm{dc}}$, we expect a dependence of $\Delta H_{sw}^{TST}$ which is quadratic in $I_{\textrm{ac}}$, since $\nabla T=A_{1}I^{2}$ with $A_{1}=\frac{\rho d}{2\kappa \pi^{2}r^{4}}$ according to Eq.~(\ref{Joule}). Taking the resistivity and Seebeck coefficients for the electrodeposited layers from earlier work~\cite{Gravier}, we can fit our data (Fig.~\ref{fig3}) assuming that the actual diameter of the nanowire is 37~nm, instead of the 50~nm average pore size of the membrane.  Such a value is realistic for an individually contacted nanowire in the very large number of them grown simultaneously.

Considering $\tau$ as a function of $\mathbf{j_{m}}$ and using Eqs.~(\ref{V2f}) and (\ref{jm}) with $I=I_{\textrm{dc}}+I_{\textrm{ac}}$, a little algebra yields:

\begin{align}\label{V2f_full}
V^{2f}_{peak}= &\frac{dR}{d\tau}\frac{d\tau}{dj_{m}}2c\left(\frac{\rho}{\pi r^{2}}+3S_{eff}A_{1}I_{\textrm{dc}}\right)I_{\textrm{ac}}^{2}+I_{\textrm{dc}}\frac{dR}{dT}\Delta T^{2f}.
\end{align}

In Fig.~\ref{fig4}, we show the calculation of Eq.~(\ref{V2f_full}) using the same values for all parameters as in the calculation of Fig.~\ref{fig3}, including the radius. The small difference between the measured data and the calculation comes from the contribution to $V^{2f}$ of the temperature dependence of the resistance (3rd term of Eq.~(\ref{V2f_full})). This has been studied extensively by what we called thermo-galvanic voltage (TGV) measurements~\cite{Granville_Ansermet}.  This technique measures $I_{\textrm{dc}}\frac{dR}{dT}\Delta T$ only. In spin valves the value is known to be a fraction of 1~$\mu V$ when $I_{\textrm{dc}}$ is 100~$\mu A$. Thus we find in Fig.~\ref{fig4} a difference between the data and the prediction of Eq.~(\ref{V2f_full}) which is of the right size for a TGV contribution.

In summary, we relied on the Joule heating in a spin valve embedded in a nanowire to produce a large local temperature gradient. We find two effects of the heat current produced in this way: a change in switching field as a function of the applied ac current, and a change as a function of the dc current of the amplitude of the peak in the second harmonic response. We rule out the spurious effects of a fixed temperature rise in the spin valve and an ac spin transfer torque by verifying that a symmetric spin valve is insensitive to a change in applied ac current. Our results provide evidence for a thermal spin transfer torque associated with the heat current. The effect of the heat current on magnetization is interpreted as the coupling of heat, charge and spin currents, as described by the thermodynamics of transport given by the Onsager reciprocity relations. This provides a reasonable estimate for the observed heat-current-driven thermal spin torque relative to the spin transfer torque.


\begin{thebibliography}{18}
\expandafter\ifx\csname natexlab\endcsname\relax\def\natexlab#1{#1}\fi
\expandafter\ifx\csname bibnamefont\endcsname\relax
  \def\bibnamefont#1{#1}\fi
\expandafter\ifx\csname bibfnamefont\endcsname\relax
  \def\bibfnamefont#1{#1}\fi
\expandafter\ifx\csname citenamefont\endcsname\relax
  \def\citenamefont#1{#1}\fi
\expandafter\ifx\csname url\endcsname\relax
  \def\url#1{\texttt{#1}}\fi
\expandafter\ifx\csname urlprefix\endcsname\relax\def\urlprefix{URL }\fi
\providecommand{\bibinfo}[2]{#2}
\providecommand{\eprint}[2][]{\url{#2}}

\bibitem[{\citenamefont{Slonczewski et~al.}(1997)\citenamefont{Slonczewski}}]{Slonczewski}
 \bibinfo{author}{\bibfnamefont{J.~C.}~\bibnamefont{Slonczewski}},
   \bibinfo{patent}{U.S.~Patent~No.~5695864} (\bibinfo{year}{1997});
    \bibinfo{author}{\bibfnamefont{J.~Z.}~\bibnamefont{Sun}},
    \bibinfo{patent}{U.S.~Patent~No.~6130814} (\bibinfo{year}{1998}).

\bibitem[{\citenamefont{Hatami et~al.}(2009)\citenamefont{Hatami, Bauer, Zhang,
  and Kelly}}]{Hatami_Bauer2}
\bibinfo{author}{\bibfnamefont{M.}~\bibnamefont{Hatami}},
  \bibinfo{author}{\bibfnamefont{G.~E.~W.} \bibnamefont{Bauer}},
  \bibinfo{author}{\bibfnamefont{Q.}~\bibnamefont{Zhang}}, \bibnamefont{and}
  \bibinfo{author}{\bibfnamefont{P.~J.} \bibnamefont{Kelly}},
  \bibinfo{journal}{Phys. Rev. B} \textbf{\bibinfo{volume}{79}},
  \bibinfo{pages}{174426} (\bibinfo{year}{2009}).

\bibitem[{\citenamefont{Katayama-Yohida
  et~al.}(2007)\citenamefont{Katayama-Yohida, Fukushima, Dinh, and
  Sato}}]{Katayama}
\bibinfo{author}{\bibfnamefont{H.}~\bibnamefont{Katayama-Yoshida}},
  \bibinfo{author}{\bibfnamefont{T.}~\bibnamefont{Fukushima}},
  \bibinfo{author}{\bibfnamefont{V.~A.} \bibnamefont{Dinh}}, \bibnamefont{and}
  \bibinfo{author}{\bibfnamefont{K.}~\bibnamefont{Sato}},
  \bibinfo{journal}{Jpn. J. Appl. Phys.} \textbf{\bibinfo{volume}{46}},
  \bibinfo{pages}{L777} (\bibinfo{year}{2007}).

\bibitem[{\citenamefont{Gravier
  et~al.}(2006{\natexlab{a}})\citenamefont{Gravier, Fukushima, Kubota,
  Yamamoto, and Yuasa}}]{Gravier_Fukushima}
\bibinfo{author}{\bibfnamefont{L.}~\bibnamefont{Gravier}},
  \bibinfo{author}{\bibfnamefont{A.}~\bibnamefont{Fukushima}},
  \bibinfo{author}{\bibfnamefont{H.}~\bibnamefont{Kubota}},
  \bibinfo{author}{\bibfnamefont{A.}~\bibnamefont{Yamamoto}}, \bibnamefont{and}
  \bibinfo{author}{\bibfnamefont{S.}~\bibnamefont{Yuasa}},
  \bibinfo{journal}{J.Phys. D: Appl. Phys.} \textbf{\bibinfo{volume}{39}}, \bibinfo{pages}{5267}
  (\bibinfo{year}{2006}{\natexlab{a}}).

\bibitem[{\citenamefont{Uchida et~al.}(2008)\citenamefont{Uchida, Takahashi,
  Harii, Ieda, Koshibae, Ando, Maekawa, and Saitoh}}]{Uchida}
\bibinfo{author}{\bibfnamefont{K.}~\bibnamefont{Uchida}} et al. 
  \bibinfo{journal}{Nature} \textbf{\bibinfo{volume}{455}},
  \bibinfo{pages}{778} (\bibinfo{year}{2008}).

\bibitem[{\citenamefont{Hatami et~al.}(2007)\citenamefont{Hatami, Bauer, Zhang,
  and Kelly}}]{Hatami_Bauer1}
\bibinfo{author}{\bibfnamefont{M.}~\bibnamefont{Hatami}},
  \bibinfo{author}{\bibfnamefont{G.~E.~W.} \bibnamefont{Bauer}},
  \bibinfo{author}{\bibfnamefont{Q.}~\bibnamefont{Zhang}}, \bibnamefont{and}
  \bibinfo{author}{\bibfnamefont{P.~J.} \bibnamefont{Kelly}},
  \bibinfo{journal}{Phys. Rev. Lett.} \textbf{\bibinfo{volume}{99}},
  \bibinfo{pages}{066603} (\bibinfo{year}{2007}).

\bibitem[{\citenamefont{Heikkil\"{a} et~al.}()\citenamefont{Heikkil\"{a},
  Hatami, and Bauer}}]{Heikkila}
\bibinfo{author}{\bibfnamefont{T.~T.} \bibnamefont{Heikkil\"{a}}},
  \bibinfo{author}{\bibfnamefont{M.}~\bibnamefont{Hatami}}, \bibnamefont{and}
  \bibinfo{author}{\bibfnamefont{G.~E.~W.} \bibnamefont{Bauer}},
  \bibinfo{journal}{Phys. Rev. B}. \textbf{\bibinfo{volume}{81}},
  \bibinfo{pages}{100408(R)} (\bibinfo{year}{2010}).

\bibitem[{\citenamefont{Yuan et~al.}(2009)\citenamefont{Yuan, Weng, and
  Xia}}]{Yuan}
\bibinfo{author}{\bibfnamefont{Z.}~\bibnamefont{Yuan}},
  \bibinfo{author}{\bibfnamefont{S.}~\bibnamefont{Weng}}, \bibnamefont{and}
  \bibinfo{author}{\bibfnamefont{K.}~\bibnamefont{Xia}},
  \bibinfo{journal}{Solid State Commun.} \textbf{\bibinfo{volume}{150}},
  \bibinfo{pages}{548} (\bibinfo{year}{2010}).

\bibitem[{\citenamefont{Bauer et~al.}()\citenamefont{Bauer, Bretzel, Brataas,
  and Tserkovnyak}}]{Bauer}
\bibinfo{author}{\bibfnamefont{G.~E.~W.} \bibnamefont{Bauer}},
  \bibinfo{author}{\bibfnamefont{S.}~\bibnamefont{Bretzel}},
  \bibinfo{author}{\bibfnamefont{A.}~\bibnamefont{Brataas}}, \bibnamefont{and}
  \bibinfo{author}{\bibfnamefont{Y.}~\bibnamefont{Tserkovnyak}},
  \bibinfo{journal}{Phys. Rev. B}. \textbf{\bibinfo{volume}{81}},
  \bibinfo{pages}{024427} (\bibinfo{year}{2010}).

\bibitem[{\citenamefont{Kovalev and Tserkovnyak}()}]{Kovalev2}
\bibinfo{author}{\bibfnamefont{A.~A.} \bibnamefont{Kovalev}} \bibnamefont{and}
  \bibinfo{author}{\bibfnamefont{Y.}~\bibnamefont{Tserkovnyak}},
  \bibinfo{journal}{Solid State Commun.} \textbf{\bibinfo{volume}{150}},
  \bibinfo{pages}{500} (\bibinfo{year}{2010}).

\bibitem[{\citenamefont{Kovalev and Tserkovnyak}(2009)}]{Kovalev1}
\bibinfo{author}{\bibfnamefont{A.~A.} \bibnamefont{Kovalev}} \bibnamefont{and}
  \bibinfo{author}{\bibfnamefont{Y.}~\bibnamefont{Tserkovnyak}},
  \bibinfo{journal}{Phys. Rev. B} \textbf{\bibinfo{volume}{80}},
  \bibinfo{pages}{100408} (\bibinfo{year}{2009}).

\bibitem[{\citenamefont{Biziere}(2009)}]{Biziere}
\bibinfo{author}{\bibfnamefont{N.} \bibnamefont{Biziere}},
  \bibinfo{author}{\bibfnamefont{E.}~\bibnamefont{Mur\`{e}}}, \bibnamefont{and}
  \bibinfo{author}{\bibfnamefont{J.-Ph.}~\bibnamefont{Ansermet}},
  \bibinfo{journal}{Phys. Rev. B} \textbf{\bibinfo{volume}{79}},
  \bibinfo{pages}{012404} (\bibinfo{year}{2009}).

\bibitem[{\citenamefont{Chlebny et~al.}(1993)\citenamefont{Chlebny, Doudin, and
  Ansermet}}]{Chlebny_Doucin}
\bibinfo{author}{\bibfnamefont{I.}~\bibnamefont{Chlebny}},
  \bibinfo{author}{\bibfnamefont{B.}~\bibnamefont{Doudin}}, \bibnamefont{and}
  \bibinfo{author}{\bibfnamefont{J.-Ph.} \bibnamefont{Ansermet}},
  \bibinfo{journal}{Nanostructured Mat.} \textbf{\bibinfo{volume}{2}},
  \bibinfo{pages}{637} (\bibinfo{year}{1993}).

\bibitem[{\citenamefont{Gravier
  et~al.}(2006{\natexlab{b}})\citenamefont{Gravier, Serrano-Guisan, Reuse, and
  Ansermet}}]{Gravier}
\bibinfo{author}{\bibfnamefont{L.}~\bibnamefont{Gravier}},
  \bibinfo{author}{\bibfnamefont{S.}~\bibnamefont{Serrano-Guisan}},
  \bibinfo{author}{\bibfnamefont{F.}~\bibnamefont{Reuse}}, \bibnamefont{and}
  \bibinfo{author}{\bibfnamefont{J.-Ph.} \bibnamefont{Ansermet}},
  \bibinfo{journal}{Phys. Rev. B} \textbf{\bibinfo{volume}{73}},
  \bibinfo{pages}{024419} (\bibinfo{year}{2006}{\natexlab{b}}).

\bibitem[{\citenamefont{Yu et~al.}(2009)\citenamefont{Yu, Dubois, Granville,
  Yu, and Ansermet}}]{Yu_Dubois}
\bibinfo{author}{\bibfnamefont{H.}~\bibnamefont{Yu}},
  \bibinfo{author}{\bibfnamefont{J.}~\bibnamefont{Dubois}},
  \bibinfo{author}{\bibfnamefont{S.}~\bibnamefont{Granville}},
  \bibinfo{author}{\bibfnamefont{D.~P.} \bibnamefont{Yu}}, \bibnamefont{and}
  \bibinfo{author}{\bibfnamefont{J.-Ph.} \bibnamefont{Ansermet}},
  \bibinfo{journal}{J. Phys. D: Appl. Phys.} \textbf{\bibinfo{volume}{42}},
  \bibinfo{pages}{175004} (\bibinfo{year}{2009}).

\bibitem[{\citenamefont{Tulapurkar et~al.}()\citenamefont{Tulapurkar, Suzuki, Fukushima, Kubota, Maehara, Tsunekawa, Djayaprawira, Watanabe, and Yuasa}}]{Yuasa}
\bibinfo{author}{\bibfnamefont{A.~A.}~\bibnamefont{Tulapurkar}} et al. 
  \bibinfo{journal}{Nature} \textbf{\bibinfo{volume}{438}},
  \bibinfo{pages}{339} (\bibinfo{year}{2005}).

\bibitem[{\citenamefont{Sankey et~al.}()\citenamefont{Sankey, Braganca, Garcia, Krivorotov, Buhrmann, and Ralph}}]{Ralph}
\bibinfo{author}{\bibfnamefont{J.~C.}~\bibnamefont{Sankey}} et al. 
  \bibinfo{journal}{Phys. Rev. Lett.} \textbf{\bibinfo{volume}{96}},
  \bibinfo{pages}{227601} (\bibinfo{year}{2006}).

\bibitem[{\citenamefont{Johnson and Silsbee}(1985)}]{Johnson_Silsbee}
\bibinfo{author}{\bibfnamefont{M.}~\bibnamefont{Johnson}} \bibnamefont{and}
  \bibinfo{author}{\bibfnamefont{R.~H.} \bibnamefont{Silsbee}},
  \bibinfo{journal}{Phys. Rev. Lett.} \textbf{\bibinfo{volume}{55}},
  \bibinfo{pages}{1790} (\bibinfo{year}{1985}).

\bibitem[{\citenamefont{Johnson and Byers}(2003)}]{Johnson_Byers}
\bibinfo{author}{\bibfnamefont{M.}~\bibnamefont{Johnson}} \bibnamefont{and}
  \bibinfo{author}{\bibfnamefont{J.}~\bibnamefont{Byers}},
  \bibinfo{journal}{Phys. Rev. B} \textbf{\bibinfo{volume}{67}},
  \bibinfo{pages}{125112} (\bibinfo{year}{2003}).

\bibitem[{\citenamefont{Wegrowe et~al.}(2007)\citenamefont{Wegrowe, Ciornei,
  and Drouhin}}]{Wegrowe}
\bibinfo{author}{\bibfnamefont{J.-E.} \bibnamefont{Wegrowe}},
  \bibinfo{author}{\bibfnamefont{M.~C.} \bibnamefont{Ciornei}},
  \bibnamefont{and} \bibinfo{author}{\bibfnamefont{H.-J.}
  \bibnamefont{Drouhin}}, \bibinfo{journal}{J. Phys.: Condens. Matter}
  \textbf{\bibinfo{volume}{19}}, \bibinfo{pages}{165213}
  (\bibinfo{year}{2007}).

\bibitem[{\citenamefont{Dubois and Ansermet}(2008)}]{Dubois_Ansermet}
\bibinfo{author}{\bibfnamefont{J.}~\bibnamefont{Dubois}} \bibnamefont{and}
  \bibinfo{author}{\bibfnamefont{J.-Ph.} \bibnamefont{Ansermet}},
  \bibinfo{journal}{Phys. Rev. B} \textbf{\bibinfo{volume}{78}},
  \bibinfo{pages}{184430} (\bibinfo{year}{2008}).

\bibitem[{\citenamefont{Granville et~al.}()\citenamefont{Granville, Yu, Dubois,
  Gravier, and Ansermet}}]{Granville_Ansermet}
\bibinfo{author}{\bibfnamefont{S.}~\bibnamefont{Granville}},
  \bibinfo{author}{\bibfnamefont{H.}~\bibnamefont{Yu}},
  \bibinfo{author}{\bibfnamefont{J.}~\bibnamefont{Dubois}},
  \bibinfo{author}{\bibfnamefont{L.}~\bibnamefont{Gravier}}, \bibnamefont{and}
  \bibinfo{author}{\bibfnamefont{J.-Ph.} \bibnamefont{Ansermet}},
  \bibinfo{journal}{J. Magn. Magn. Mater.} \textbf{\bibinfo{volume}{322}},
  \bibinfo{pages}{1464} (\bibinfo{year}{2010}).

\end{thebibliography}
\end{document}